\documentclass[seceq,preprint]{ptptex}

\usepackage{graphicx}
\notypesetlogo                       %comment in if to eliminate PTPTeX 
\preprintnumber[3cm]{%<-- [..]: optional width of preprint # column.
SAGA-HE-250}
\title{%        %You can use \\ for explicit line-break.
Analytic Continuation in  Two-color Finite Density QCD\\
 and \\
   Chiral Random Matrix Model 
}

\author{%       %Use \scshape for the family name.
Yasuhiko \textsc{Shinno}\footnote{E-mail: shinno@libe.nara-k.ac.jp}  and Hiroshi \textsc{Yoneyama}$^{\ddagger}$\footnote{E-mail: yoneyama@cc.saga-u.ac.jp}%
}

\inst{%     %Affiliation, neglected when [addenda] or [errata].
Nara National Colledge of Technology, Yamatokoriyama 639-1080,  Japan\\
Department of Physics, Saga University, Saga 840-8502, Japan$^{\ddagger}$
}

\abst{%         %This abstract is neglected when [addenda] or [errata].
Two-color finite density QCD  is free from the sign problem, and it is thus regarded as a good model to check the validity of the analytic continuation method. We study the method in terms of  the corresponding chiral random matrix model.  It is found that at temperatures slightly higher than the pseudo critical temperature, the ratio type of   extrapolated function works  well in accordance with the results of the Monte Carlo simulations. 
}

\begin{document}

\maketitle

\section{Introduction}
\label{sec:introduction}
Determining the phase structure of QCD in the chemical potential ($\mu$)- temperature ($T$)  plane is an 
important  issue to understand  properties of dynamics of QCD.  
Lattice gauge theory is one of  promising tools to  study non perturbative properties of QCD. At finite values of $\mu$,  however, it suffers from notorious sign problem. 
Fermion  determinant takes  complex values and this invalidates importance sampling in Monte Carlo (MC) simulations. 
In order to circumvent the problem, various methods  such as Taylor series method,~\cite{rf:All}\cite{rf:GG} \ reweighting method~\cite{rf:FK} \ etc. are attempted.  
Among such methods is analytic continuation~\cite{rf:dFP,rf:DL,rf:GP,rf:ADGL,rf:CL,rf:WLC} \  that we pay attention to in the present paper.  In this method, one calculates some  quantities in imaginary chemical potential region, fit the data by an appropriate function and extrapolate it   to the  real $\mu$ region.    
Such an extrapolation,  in general,  requires careful assessment of extrapolated function.  
In order to explicitly check the validity of   extrapolated functions,  Cea et.al.~\cite{rf:CCDP} \cite{rf:CCDP2} \  studied two-color QCD,~\cite{rf:N,rf:DFK,rf:DMW,rf:KST,rf:K,rf:KTS,rf:MNN,rf:SSS,rf:KSHM,rf:KTS2,rf:STV,rf:WLS,rf:RW2,rf:NFH,rf:AEL,rf:FI} \ which  is free from the sign problem. They  calculated  several quantities such as chiral condensate and Polyakov loop in the imaginary $\mu$ region,  and compared these  extrapolations with  the results of direct  computations  in the real $\mu$ region.  
 Based on what they studied, they made useful  suggestions  for applying the analytic continuation method to  three-color QCD.   In the current stage,   however, it is a hard task  to check their statements by using MC simulations because of the sign problem. It would then  be useful  to study these properties by adopting some other methods. In the present paper we use chiral random matrix theory  (RMT) model.~\cite{rf:VW}   \par
Chiral RMT is equivalent to QCD in the $\epsilon$ regime. 
This microscopic equivalence is made use of for calculating  low energy constants such as pion decay constant.  From a mean field point of view, on the other hand, it is used to  study phase diagram in the $\mu-T$ plane and has predicted  relevant properties.~\cite{rf:HJSSV}  \ Here, taking   the latter stand point we study  analytic continuations of two-color QCD  in terms of    the corresponding chiral RMT  by investigating  transitive regions  from imaginary to real $\mu$ region at finite $T$.   Based on this study, we shall investigate three-color QCD, the results of which   will be reported in a  forthcoming paper.\cite{rf:SY}  \   In the present paper, we adopt a model proposed by Klein, Toublan and Verbaarschot.~\cite{rf:KTV2,rf:KTV}  \ Since we are concerned about analytic continuation properties, we study the phase structure in the  imaginary $\mu$ region, mainly in the vicinity of $\mu=0$. 
Although the phase structure in the real    $\mu$ region is investigated in Ref.~\citen{rf:KTV},  we present a part of these  results associated with analytic continuation.  \par
It is found in  MC studies~\cite{rf:CCDP} \cite{rf:CCDP2} \ that  suitable   fitting function  
depends  on the region of  temperatures, possibly reflecting the difference of involved physics. 
In particular, at moderately  high temperature region, the ratio type fitting function works well,  and it is suggested to use it in the study of  three-color QCD.  
In the present paper,  we study the  phase structure of the chiral RMT model  in the real and  imaginary $\mu$ regions in the vicinity of $\mu=0$.  Based on the results, we  divide the temperature  into three regions, $T<T_D$ (region I),   $T_D<T<T_{co}$ (region II) and  $T_{co}<T$ (region III), where $T_D$ is the maximal temperature  of the  diquark condensate phase  in the real $\mu$ region,   and $T_{co}$ is temperature of cross over at $\mu=0$.  We  study the analytic continuation of the chiral condensate and the cross over line.  As fitting functions, we use the polynomial types as well as the ratio type.  In region III, we found that the polynomial fits  show slow convergence, while the ratio type fits work well,  in accordance with the results of MC simulation studies.~\cite{rf:CCDP} \ This is the main results of the present paper.  \par
 The adopted RMT  model  incorporates   a temperature effect  only through  the lowest Matsubara frequencies.    This approximation  is expected to be poor   in the low temperature region.  
 In order  to check if the results stated above concerning the analytic continuation is sensitive to this approximation, we also consider  an extended RMT model~\cite{rf:VJ}, \ which incorporates all the Matsubara frequencies.  
 We study the phase structure of this model  in the real and  imaginary $\mu$ regions. In the real $\mu$ region,  we make it clear   the relationship of temperature with that in the original RMT model.  In the region III,  we found that the ratio type of  function fits well the behaviors of the chiral condensate  in this model, too.    
 %
 %Since space-time structure is not emerged explicitly in the RMT. 
  \par
  In the following section, we present the formulation of  two-color QCD and its  corresponding chiral RMT model.  We  briefly review its phase structure  in the real $\mu$ region and study the phase structure  in the  imaginary $\mu$ region.  We also present the results of the phase structure in the presence  of a diquark source rather than quark mass.  In section 3, we study analytic continuation of  the chiral condensates  and the pseudo critical line.  In section 4, we  discuss an effect of the higher Matsubara frequencies.
   Summary is presented in section 5.

\section{Phase structure}
\label{sec:sec2}
\subsection{Two-color QCD and  RMT}
In this subsection, let us  briefly remind us of the chiral  Lagrangian study and   chiral random matrix model in  two-color QCD for a completeness.  Two-color QCD is  characterized by pseudo reality  
\begin{equation}
\tau_2\tau_a\tau_2=-\tau_a^{*},
\end{equation}
where $\tau_a$ are Pauli matrices. As a consequence,  the    Lagrangian with massless  $N_f$ flavor quarks  possesses enlarged  SU($2N_f$) flavor symmetry  rather than  
SU($N_f$)$\times$SU($N_f$) symmetry.  An introduction of the  chemical potential induces symmetry breaking  
SU($2N_f$)$\rightarrow$ SU($N_f$)$\times$ SU($N_f$)$\times$ U(1)$_B$. For non zero quark mass $m\neq 0$, 
SU($2N_f$) turns to  Sp($2N_f$),  and further Sp($2N_f$) $\rightarrow$  SU$(N_f)_V$$\times$ U(1)$_B$ due to a  $\mu$ term. 
Some time ago,  Kogut et. al.~\cite{rf:KST,rf:K} \ studied the model  based on the  chiral Lagrangian.
The static terms to    lowest order  in the  momentum expansion of the effective Lagrangian are given by   
\begin{equation}
\mathcal{L}_{\rm chiral}^{\rm static}=-F^2m_{\pi}^2{\rm Re}{\rm Tr }\left(\Sigma {\hat M}\right)
-F^2\mu^2{\rm Tr }\left(\Sigma B^{T}\Sigma^{\dagger} B +BB \right),
\end{equation}
where $\Sigma$ is a $2N_f\times 2N_f$ unitary antisymmetric  matrix. The  mass matrix ${\hat M}$  and baryon charge matrix $B$ 
are given by
\begin{equation}
{\hat M} =\left( \begin{array}{cc}
  0 & 1 \\
  -1 & 0
\end{array} \right), \quad B=\left( \begin{array}{cc}
  1 & 0 \\
  0 & -1
\end{array} \right),
\end{equation}
where $1$ is $N_f\times N_f$ unit matrix.  
The constant $F$ is a low energy constant,  and pion mass is given in terms of chiral condensate $\langle {\bar \psi}\psi \rangle_0$ at $\mu=0$ by  
$m_{\pi}^2=m\langle {\bar \psi}\psi \rangle_0/(2N_f F^2)$. 
The Lagrangian  can  be minimized by  using the following form 
\begin{equation}
 \Sigma=\Sigma_{\sigma} \cos \alpha +\Sigma_{\Delta}\sin \alpha,  
\end{equation}
where 
\begin{equation}
\Sigma_{\sigma} =\left( \begin{array}{cc}
  0 & -1 \\
  1 & 0
\end{array} \right), \quad \Sigma_{\Delta}=\left( \begin{array}{cc}
  iI & 0 \\
  0 & iI
\end{array} \right), \quad I=\left( \begin{array}{cc}
  0 & -1 \\
  1 & 0
\end{array} \right).
\end{equation}
For $\mu\leq m_{\pi}/2$, the chiral condensate dominates, while  for $\mu>m_{\pi}/2$ a rotation of condensates to the diquark condensate occurs as
\begin{eqnarray}
\Sigma=\left\{ \begin{array}{ll}
\Sigma_{\sigma} &  ( \mu\leq m_{\pi}/2) \\
 \Sigma_{\sigma} \cos \alpha_0 + \Sigma_{\Delta}\sin \alpha_0
 &  (\mu>m_{\pi}/2), 
\end{array} \right.
\label{eq:rotate}
\end{eqnarray}
where  $ \cos \alpha_0 = m_{\pi}^2/(4\mu^2)$.\par
%%%%%%%%%%%%%%%%%%%%%%%%%%%%%%%%%%%
The QCD partition function with two flavors 
\begin{equation}
Z_{\rm QCD}=\langle \prod_{f=1}^{2} \det (D+m_f +\mu_f\gamma_0)\rangle 
\end{equation}
is replaced by random matrix partition function
\begin{equation}
Z_{\rm RMT}=\int \mathcal{D}W \exp[-\frac{n}{2}G^2{\rm Tr} W^{T}W ]\det {\hat D},
\end{equation}
where due to the pseudo reality,  $W$ is represented by a  real  $n\times n$ matrix,  and the probability distribution of the matrix elements is Gaussian. The matrix $ {\hat D}$ is given by~\cite{rf:KTV}
\begin{eqnarray*}
& {\hat D} &= \\
 & &\left(\begin{array}{cccc}
m_1 & 0 & W+\omega(T) +\mu_1& 0 \\
0 &  m_2 &  0 & W+\omega(T) +\mu_2\\
-W^{T}-\omega(T)^{T} +\mu_1&  0 &  m_1  & 0\\
0 &  -W^{T}-\omega(T)^{T} +\mu_2&  0 & m_2
\end{array} \right).
\end{eqnarray*}
As a temperature effect, only the Matsubara lowest frequencies  are  incorporated in the form  
\begin{eqnarray}
\omega(T)=\left( \begin{array}{cc} 0 & T \\ -T & 0 
\end{array} \right).
\end{eqnarray}
In the present paper, we consider symmetric case $\mu_1=\mu_2$, and  $m_1=m_2$. \\
Standard manipulations lead to a partition function represented by complex $2N_f\times 2N_f$ $(N_f=2)$ matrix
 $A$~\cite{rf:KTV}
 \begin{eqnarray}
Z_{\rm RMT}&=&\int \mathcal{D}A \exp[-\mathcal{L} ], \label{eq:zrmt}\\
\mathcal{L}&=&\frac{n}{2}G^2{\rm Tr} AA^{\dagger}-\frac{n}{4}\log\det Q_+ Q_-,
\end{eqnarray}
where 
\begin{eqnarray}
 Q_{\pm} =\left( \begin{array}{cc} A^{\dagger}+M  &  \mu B \pm iT \\ - \mu B \pm iT & A+M^{\dagger}
\end{array} \right), 
\quad 
 M =\left( \begin{array}{cccc}
  0   & m  & 0 &0  \\ 
-m  & 0 & 0  & 0\\
0 & 0 & 0 & m\\
0 & 0  &-m  & 0
\end{array} \right)
\label{eq:QM}
\end{eqnarray}
%}
%
 and $B={\rm diag}(1,-1,1,-1)$. \\
An ansatz for $A$
\begin{eqnarray}
 A =\left( \begin{array}{cccc}
  0   & -\sigma & -i\Delta &0  \\ 
\sigma & 0 & 0  &  -i\Delta\\
i\Delta & 0 & 0 & -\sigma\\
0 & i\Delta & \sigma & 0
\end{array} \right)
\end{eqnarray}
leads to an effective potential:
\begin{eqnarray}
\frac{1}{n}\mathcal{L}&=&2G^2(\sigma^2+\Delta^2)\nonumber-\ln D,\label{eq:action}\\
D&=& 
\left[\left(\sigma + m+\mu \right)^2+\Delta^2 +T^2 \right]\left[\left(\sigma + m-\mu \right)^2+\Delta^2 +T^2 \right]. 
\end{eqnarray}

%%%%%%%
\subsection{saddle point equation}
In order to obtain the most dominant contribution in Eq.(\ref{eq:zrmt}) in the $n\rightarrow \infty$ limit,    we solve the saddle point equations  given by
\begin{equation}
\frac{\partial \mathcal{L}}{\partial \sigma}=0, \quad \frac{\partial \mathcal{L}}{\partial \Delta}=0.
\end{equation}
This yields 
\begin{eqnarray}
 \sigma G^2D-\left(\left(\sigma+m\right)^2-\mu^2+\Delta^2+T^2\right)(\sigma+m)=0\label{eq:sdp1}, \\
\Delta \left[G^2D-\left((\sigma+m)^2+\mu^2+\Delta^2+T^2\right)\right]=0,
\label{eq:sdp2}
\end{eqnarray}
where $D$  reads 
\begin{eqnarray}
D&=&\left[\left(\sigma+m\right)^2+\Delta^2+\mu^2\right]^2-4\mu^2\left(\sigma+m\right)^2\nonumber\\
&+&2\left[\left(\sigma+m\right)^2+\Delta^2+\mu^2\right]T^2+T^4.
\label{eq:d}
\end{eqnarray}
\\
Saddle point equations  in the imaginary $\mu$ region are given by replacing $\mu$ by $i\phi $  in Eq's.~(\ref{eq:sdp1}), (\ref{eq:sdp2}) and  (\ref{eq:d}). 
In the following subsections, the phase structure in the imaginary $\mu$ region is studied.  In order to  understand  the following  analysis of the analytic continuation, we also discuss the phase  in the real $\mu$ region. Although the latter is studied in detail in  Ref. \citen{rf:KTV},   we present a part of it in a self-contained manner.  It should be noted that the RMT model does not show a periodicity in the imaginary $\mu$,   caused by the center symmetry of the gauge group (Roberge-Weiss symmetry~\cite{rf:RW} \ ).   This point is discussed   in summary  in connection with the analytic continuation. 
%
%%%%%%%%%%%%%%%%%%%%%%%%%%%%%%%%%%%%%
%%%%%%%%%%%%%%%%%%%%%%%%%%%%%%%%%%%%%
\subsubsection{$m=0$}
In the following section, we shall discuss the analytic continuation in the case of non zero quark mass.  In order to have a better  understanding  of   the phase structure in the massive case, we also discuss the phase structure in  the massless case. 
For $m=0$, the saddle point equations are analytically solved.
For $T=0$, the saddle point equations become
\begin{eqnarray}
\left\{ \begin{array}{ll}
\sigma=0, & G^2D=\sigma^2-\mu^2+\Delta^2,\\
\Delta=0, & G^2D=\sigma^2+\mu^2+\Delta^2, 
 \end{array}\right.
\end{eqnarray}
\begin{eqnarray}
D=(\sigma^2-\mu^2+\Delta^2)^2+4\Delta^2\mu^2.
\end{eqnarray}
This is classified into  the following four  cases:(a) $\sigma=0, \Delta=0$, (b) $\sigma=0, \Delta\neq 0$,  (c) $\sigma\neq 0, \Delta=0$, (d) $\sigma\neq 0, \Delta\neq 0$. 
 Correspondingly, the effective potential reads
\begin{eqnarray}
\Omega&\equiv& \frac{1}{n}\mathcal{L}^{(a)}=-2\ln \mu^2,\nonumber\\
\Omega^{(\Delta)}&\equiv& \frac{1}{n}\mathcal{L}^{(b)}=2(1+\ln G^2-G^2\mu^2) \quad (\Delta^2+\mu^2=1/G^2),\nonumber\\
\Omega^{(\sigma)}&\equiv& \frac{1}{n}\mathcal{L}^{(c)}=2(1+\ln G^2+G^2\mu^2) \quad (\sigma^2-\mu^2=1/G^2),\nonumber\\
\Omega^{(\Delta\sigma)}&\equiv& \frac{1}{n}\mathcal{L}^{(d)}=2(1+\ln G^2) \quad {\rm at \ \mu=0 \ only}.
\label{eq:effpot1}
\end{eqnarray}
For $0\leq \mu <\mu_c$, $\Omega^{(\Delta)}$ gives the lowest values among these $\Omega$, i.e., 
diquak condensation occurs,  where $\mu_c$ is determined so that  $2(1-G^2\mu^2)+\ln (G^2\mu^2)^2=0 \ (\Omega^{(\Delta)}=\Omega)$ is satisfied.   For  $\mu >\mu_c$, symmetric phase realizes. 
For imaginary $\mu=i\phi$,  one can easily see in  Eq.~(\ref{eq:effpot1}) that   minimum  is found  just by  an interchange of $\sigma$ and $\Delta$, i.e., the roles of $\sigma$ and $\Delta$  are interchanged  in the real and imaginary $\mu$ regions. 
 For $0\leq \phi <\phi_c$,  chiral condensation occurs,  and for $\phi_c<\phi$, $\sigma=0, \Delta=0$. The value of  $\phi_c$ satisfies $2(1-G^2\phi_c^2)+\ln (G^2\phi_c^2)^2=0 \ ({\tilde\Omega}^{(\sigma)}={\tilde \Omega})$, where ${\tilde\Omega}^{(\sigma)}\equiv \Omega^{(\sigma)}(\mu\rightarrow i\phi)$ etc..  At $\mu=0$, the system undergoes a first order phase transition.  When one crosses the origin  from real $\mu$ to imaginary $\mu$ region,
the magnitudes of  $\Delta$ in the real $\mu$ region jumps down to 0 at $\mu=0$, while  that of  $\sigma$ jumps up to a finite value from vanishing one. 
 \par
\begin{figure}
%\vspace{-10mm}
        \centerline{\includegraphics[width=15cm, height=7
cm]{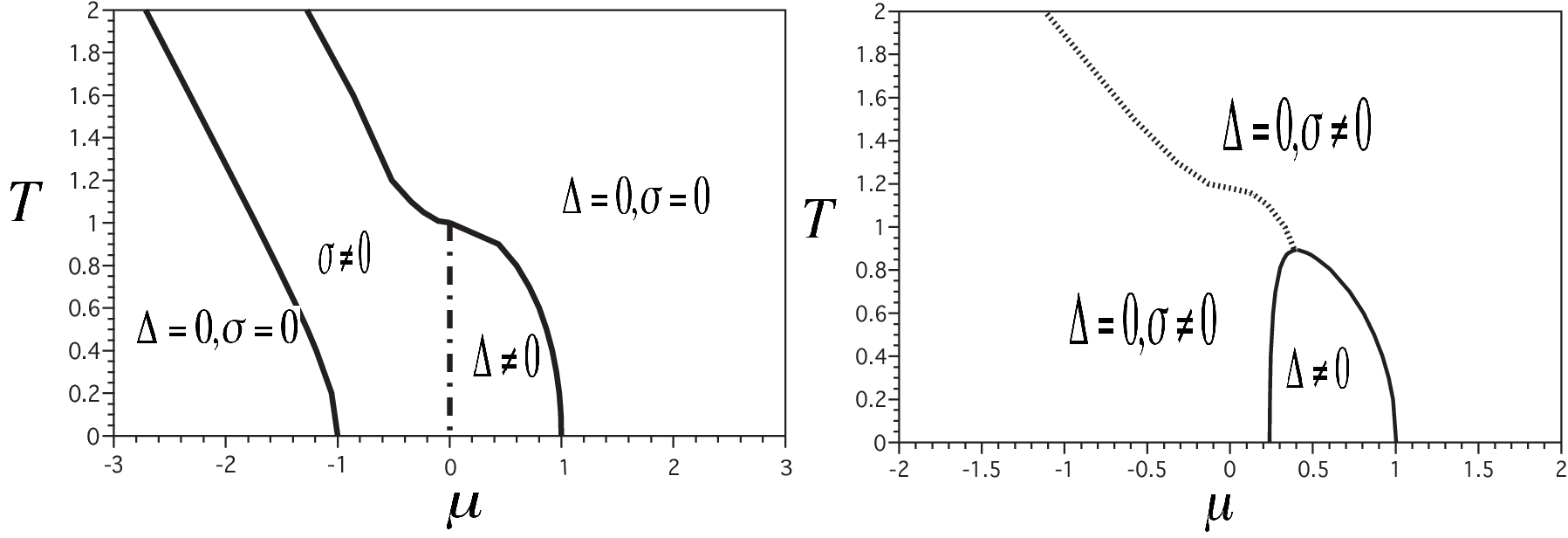}}
\caption{Phase diagram for $m=0$ (left)  and $m=0.1$ (right).  Solid lines indicate second order critical lines.  Dashed dotted  and dotted    lines  indicate  first order  critical line  and cross over, respectively.   $G=1.0$. Note that negative side of the horizontal axis  indicates the  imaginary chemical potential,  $i|\mu|$.  
 }
\label{fig:phasem0}
\end{figure}

At finite temperatures, 
this first order critical point persists up to a finite value $T_c$ along the $T$-axis as shown in the left panel in Fig.\ref{fig:phasem0}.  Temperature effect also  shifts the location of $\mu_c$ such that 
\begin{equation}
1+\ln G^2(\mu_c^2+T^2)-G^2(\mu_c^2+T^2)=0, 
\end{equation}
while $\phi_c$ is determined by  
\begin{equation}
2G^2(\phi_c^2+T^2)-\left(1+ \sqrt{1+16\phi_c^2T^2G^4}\right)=0.
\end{equation}
The second order critical line  in the real $\mu$ region starts at $\mu=\mu_c$,  passes  the endpoint  $(\mu=0, T_c)$  of the first order critical line at $\mu=0$ and  extends to the imaginary $\mu$ region  as  $T$ increases.   On the other hand, the critical line, separating symmetric and chiral condensate phases  in the imaginary $\mu$ region,  starts  at    $\phi_cG=1.0$ and  $T=0$,   and  moves toward larger values of $\phi$ as $T$ increases. These are  shown in the left panel  in Fig.\ref{fig:phasem0}. 
%%%%%%%%%%%%%%%%%
\subsubsection{$m\neq 0$}
\label{sec:sunsec}
\begin{figure}
%\vspace{-10mm}
        \centerline{\includegraphics[width=13cm, height=5
cm]{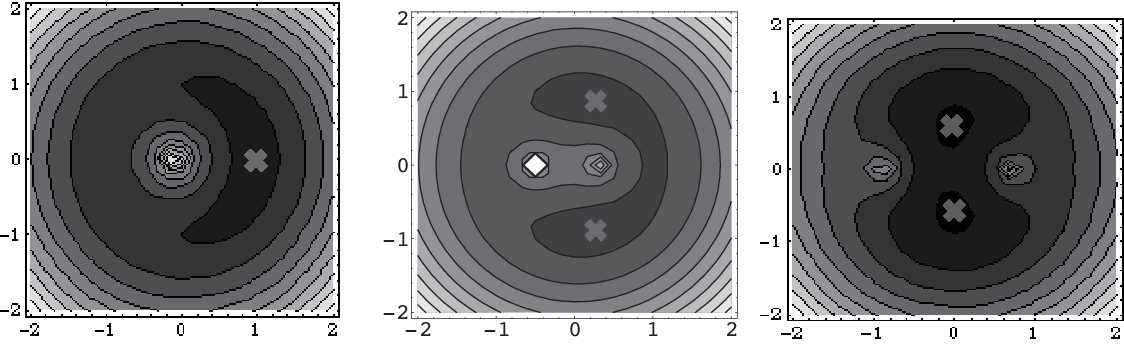}}
\caption{Contour plots of  the effective potential in $\sigma-\Delta$ plane  for  $m=0.1$ at  $T=0$. From left to right, $\mu=0.0, 0.4$ and 0.8.  The cross symbol indicates the location of the minima. $G=1.0$.   The horizontal (vertical) axis stands for $\sigma$ ($\Delta$). 
 }
\label{fig:efp3d}
\end{figure}

For $m\neq 0$, the effect of chiral condensate becomes stronger and the phase structure changes. 
The first order critical line, separating chiral and diquark condensate phases  at $\mu=0$ in the $m=0$ case,   changes to the second order one and shifts  towards real $\mu$ region as shown on the right panel of Fig.\ref{fig:phasem0}.  This is given as  the lowest energy solution of    
Eq's.~(\ref{eq:sdp1}) and (\ref{eq:sdp2}) for  small values of $\mu$.   At $T=0$, the phase transition  occurs  at $\mu_cG=0.26\equiv \mu_{c1} G$. 
For  $ \mu\geq \mu_{c1}$, chiral condensate rotates into diquark condensate, which is in agreement  with the chiral Lagrangian study as in Eq.~(\ref{eq:rotate}). 
Figure~\ref{fig:efp3d} indicates  contour plots of the effective potential  in the $\sigma-\Delta$ plane for  $m=0.1$ and  $T=0$ ($\mu G=0.0, 0.4$ and 0.8 from left to right).  The darker  the plot  becomes, the smaller the magnitude of  the  effective potential becomes. We see that the minimum rotates from $(\sigma\neq 0, \Delta=0)$ to $(\sigma=0, \Delta\neq 0)$ as $\mu$ increases. 
 Unlike the chiral Lagrangian,  however, diquark condensate becomes  vanishing  at $\mu G= \mu_c G (=1.0)$ at $T=0$, and  the diquark condensate phase turns to quasi symmetric  one  ($\sigma\approx O(m), \Delta=0$).  This is due  to the saturation effect coming from the limited dirac spectrum of the RMT. As temperature  increases,   diquark condensate  phase becomes narrower and terminates at $(\mu_D, T_D)$. 
 For $T \geq T_D$, thermal fluctuations become large enough to make   diquark  condensate cease.  A cross over line leaves the point $(\mu_D, T_D)$  and moves towards  $\mu=0$ as temperature increases,  where the signal of cross over  is detected by a peak of chiral susceptibility. 
 This  line  is a reminiscent  of the second order critical line in the $m=0$ case. 
In the imaginary $\mu$ region, it  further proceeds toward  larger values of $\phi$ as temperature increases.
For larger values of $\phi$, the signal of the critical line in the $m=0$ case,  emanating at $\phi_c$, becomes so  weak in the $m\neq 0$ case  that the peak of the susceptibility is very broad and  low.  In the figure, we do not show it. 
%
%%%%%%%%%%%%%%%%%%%%%%%%%%%%%
\subsection{diquak source $j_d\neq 0$}
\begin{figure}
%\vspace{-5mm}
        \centerline{\includegraphics[width=9cm, height=7
cm]{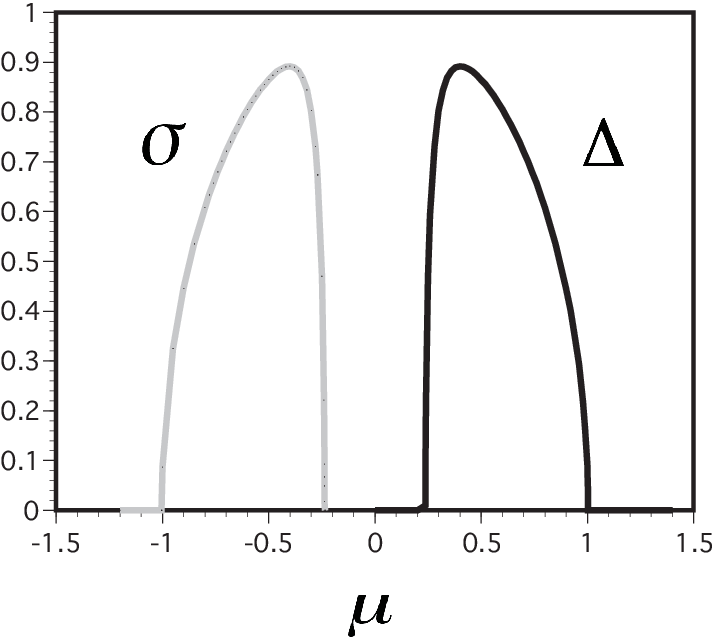}}
\caption{$\sigma$  for $(j, m)=(0.1, 0)$  and $\Delta$ for $(j, m)=(0, 0.1)$ at  $T=0$.  $G=1.0$. Negative side of the horizontal axis  indicates the  imaginary chemical potential,  $i|\mu|$.   
}
\label{fig:sigdelT0}
\end{figure}

\begin{figure}
%\vspace{-5mm}
        \centerline{\includegraphics[width=11cm, height=7
cm]{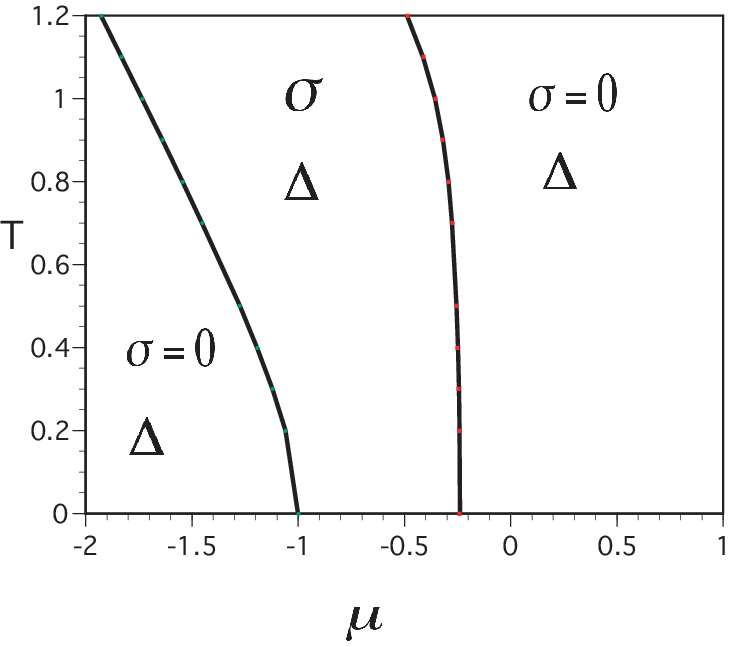}}
\caption{Critical lines   for $j=0.1$ and $m=0$. $G=1.0$.  Solid line  indicates the second order
critical line.  Negative side of the horizontal axis  indicates the  imaginary chemical potential,  $i|\mu|$.   
}
\label{fig:phasej01}
\end{figure}

So far we have discussed the phase structure in the case of non zero quark mass.  
In two-color QCD,  meson  and diquark states  belong to the same representation, and then mass and diquark source are on an equal footing. 
It is interesting,   therefore, to discuss the phase structure when  a diquak source is switched on.
 When  $m=0$,  a diquark source $j_d$ is introduced in Eq.~(\ref{eq:QM}) as 
\begin{eqnarray}
 M =\left( \begin{array}{cccc}
  0   & 0  & -ij_d &0  \\ 
0  & 0 & 0  & -ij_d\\
ij_d & 0 & 0 & 0\\
0 &  ij_d  & 0  & 0
\end{array} \right).
\end{eqnarray}
The effective potential  becomes 
\begin{eqnarray}
\frac{1}{n}\mathcal{L}&=&2G^2(\sigma^2+\Delta^2)-\ln D^{(\rm d)}\label{eq:action2}, \\
D^{(\rm d)}
&=&\left[\left(\Delta + j_d\right)^2+\sigma^2-\mu^2\right]^2+ 4\mu^2(\Delta+j_d)^2\nonumber\\
&+&2\left[\left(\Delta + j_d\right)^2+\sigma^2+\mu^2\right]T^2+T^4.
\label{eq:d2}
\end{eqnarray}
At $T=0$, the argument of the logarithmic  term in  Eq.~(\ref{eq:action2}) reads
\begin{equation}
D^{(\rm d)}=\left[\left(\Delta + j_d\right)^2+\sigma^2-\mu^2\right]^2 +4\mu^2\left(\Delta + j_d\right)^2.
\end{equation}
Comparing this with $D$ in  Eq.~(\ref{eq:d})  at $T=0$, we recognize that 
 there is a symmetry with respect to a simultaneous interchange of $j_d \leftrightarrow m$,  $\Delta \leftrightarrow\sigma$ and $\mu^2 \leftrightarrow -\mu^2$. This implies that the behavior of  $\sigma$ ($\Delta$)  in the $j_d\neq 0, m=0$ case  is a mirror image of the behavior of    $\Delta$ ($\sigma$)   in the $j_d=0, m\neq 0$ case  if one plots them as  functions of  $\mu^2$ as shown in  Fig.~\ref{fig:sigdelT0}.
The rotation of condensate thus  occurs in the imaginary $\mu$ region, from  $\Delta\neq 0$ to $\sigma\neq 0$ as $\phi$ increases from 0.\\
At  finite $T$, a term proportional to $T^2$ breaks this  symmetry  as seen in Eq.'s~(\ref{eq:d}) and  (\ref{eq:d2}). At low temperatures, such a breaking effect is small, but
at higher temperatures, it  becomes clearer.  By investigating the behaviors of $\sigma$ and $\Delta$  in the similar way to the $m\neq 0$ case, 
we obtain the critical lines   as  shown   in Fig.~\ref{fig:phasej01}.  Unlike the $m\neq 0,  j_d=0$ case, the phase with both of  $\sigma$ and $\Delta$  condensation, located now in the imaginary $\mu$ region,   does not close  itself    but stretches  to  larger temperature region. 
% %

 \section{Analytic continuation}
\label{sec:sec3}
\begin{figure}
%\vspace{-5mm}
        \centerline{\includegraphics[width=14cm, height=7
cm]{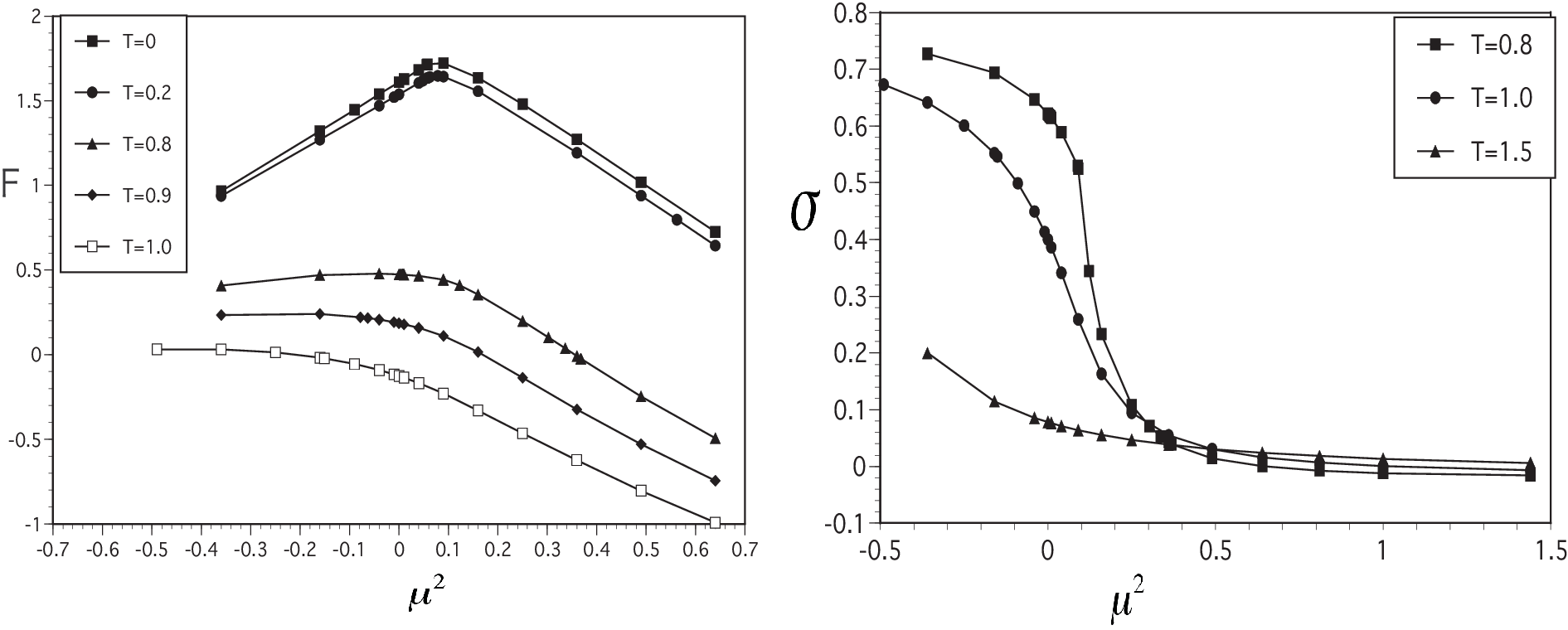}}
\caption{Free energy (left)  and chiral condensate (right) in the vicinity of $\mu=0$  for $m=0.1$ and various values of  $T$. 
 }
\label{fig:effpot-im-1}
\end{figure}

In this section, we discuss analytic continuation for $m\neq 0$.  Hereafter, G is taken to be 1.0. 
Figure~\ref{fig:effpot-im-1} indicates  free energy and chiral condensate in the real and imaginary $\mu$ regions for various temperatures for $m=0.1$.
Both of the quantities behave smoothly at $\mu=0$ for all temperatures considered here.
Based on these results, we check analytic continuation in terms of the RMT and compare the results with those of MC simulations.~\cite{rf:CCDP} \cite{rf:CCDP2}
\subsection{chiral condensate}
We use  calculated values of  $\sigma$   in the imaginary $\mu$ region  as ``data".  We take $s$ points of data in a certain range $0<\phi \leq \phi_f$,    fit them  by  an appropriate function  and extrapolate it to the real  $\mu$ region. 
Following  Cea et. al., the following two types of fitting function are  used, one being   polynomial  type and the other  ratio type. 
\begin{eqnarray*}
\sigma=\left\{ \begin{array}{ll}
 A+B\phi^2+C\phi^4+\dots  &  ({\rm polynomial } ),\\
 \frac{A+B\phi^2}{1+C\phi^2}  &  ({\rm ratio }).
\end{array} \right.
\end{eqnarray*}
For a systematic study, we check  to what extent  the outcome depends on the following points: 
\begin{enumerate} 
\item the range of used  ``data" in the imaginary $\mu$ region
\item the number of  degrees of freedom of  the data
\item the highest degree of the polynomial  in the case of polynomial  fit
\end{enumerate}
  We used  $\phi_f=0.3$, 0.5 and 0.8,  and  found that the outcome depends rather strongly  on the adopted range  $\phi_f$.   Generally speaking,   the $\phi_f=0.3$ case
   turns  out the best among  all the cases we investigated.     On the other hand,  when  we varied  the number of degrees of freedom,  d.o.f.,     from 1  to about 75, depending on data,    the results depend  hardly on the number of data  $s$.  This is shown in  Table~\ref{t:dof}   for the  $T=0.8$ and 1.5 cases.  Hereafter,  we present the results of $\phi_f=0.3$. 
\begin{table}
 \caption{Dependence of fitting parameters on the number of data for $T=0.8$ and 1.5.  The adopted range for the data is $0.0<\phi<0.3$. $\Delta \phi$ is defined by  the number of points $s$ as  $ \phi_f/s$.  }
   \label{t:dof}
\begin{center}
{\footnotesize
 \hspace*{-10mm}
 \begin{tabular}{c||cc|cc}\hline\hline
 \multicolumn{5}{c}{$f(\phi)=A+B\phi^2+C\phi^4$} \\ \hline
{}& \multicolumn{2}{c|}{$T=0.8$}&\multicolumn{2}{c}{$T=1.5$}\\ \hline
 {$\Delta\phi$}&{0.01}&{0.05}&{0.01}&{0.05}\\ \hline
 {d.o.f.}&{28}&{4}&{28}&{4}\\
 {$\chi^2/{\rm d.o.f.}$}&{$5.63814\times 10^{-9}$}&{$1.08933\times 10^{-8}$}
 &{$5.2598\times 10^{-11}$}&{$1.02758\times 10^{-10}$}\\
 {$A$}&{0.62178(2)}&{0.62178(7)}
 &{0.07803(0)}&{0.07803(0)}\\
 {$B$}&{-0.68756(169)}&{-0.68116(454)}
 &{-0.17377(16)}&{-0.17360(44)}\\
 {$C$}&{-1.67541(2050)}&{-1.6443(506)}
 &{0.32745(198)}&{0.33086(492)}\\ \hline \hline
 \multicolumn{5}{c}{$g(\phi)=A+B\phi^2+C\phi^4+D\phi^6$} \\ \hline
 {}&\multicolumn{2}{c|}{$T=0.8$}&\multicolumn{2}{c}{$T=1.5$} \\ \hline
 {$\Delta\phi$}&{0.01}&{0.05}&{0.01}&{0.05}\\ \hline
 {d.o.f.}&{27}&{3}&{27}&{3}\\
 {$\chi^2/{\rm d.o.f.}$}&{$3.18334\times 10^{-11}$}&{$5.12716\times 10^{-11}$}
 &{$3.03625\times 10^{-14}$}&{$4.60497\times 10^{-14}$}\\
 {$A$}&{0.62171(0)}&{0.62170(0)}
 &{0.07802(0)}&{0.07802(0)}\\
 {$B$}&{-0.70553(29)}&{-0.70565(74)}
 &{-0.17551(0)}&{-0.17550(2)}\\
 {$C$}&{-2.26163(849)}&{-2.26256(2152)}
 &{0.27069(2)}&{0.27073(65)}\\
 {$D$}&{-4.65104(6623)}&{-4.63419(15920)}
 &{-0.45033(205)}&{-0.45081(477)}\\ \hline \hline
 \multicolumn{5}{c}{$h(\phi)=\frac{A+B\phi^2}{1+C\phi^2}$}\\ \hline
 {}&\multicolumn{2}{c|}{$T=0.8$}&\multicolumn{2}{c}{$T=1.5$}\\ \hline
 {$\Delta\phi$}&{0.01}&{0.05}&{0.01}&{0.05} \\ \hline
 {d.o.f.}&{28}&{4}&{28}&{4}\\
 {$\chi^2/{\rm d.o.f.}$}&{$4.40095\times 10^{-11}$}&{$9.37848\times 10^{-11}$}
 &{$1.93257\times 10^{-12}$}&{$3.59884\times 10^{-12}$}\\
 {$A$}&{0.62169(0)}&{0.62169(0)}
 &{0.07802(0)}&{0.07802(0)} \\
 {$B$}&{-2.88150(288)}&{-2.88685(769)}
 &{-0.05713(15)}&{-0.05738(37)} \\
 {$C$}&{-3.49352(434)}&{-3.50174(1153)}
 &{1.52002(163)}&{1.51722(399)} \\ \hline
 \end{tabular}
 }
 \end{center}

\end{table}

\begin{figure}
\vspace{5mm}
        \centerline{\includegraphics[width=14cm, height=7.5
cm]{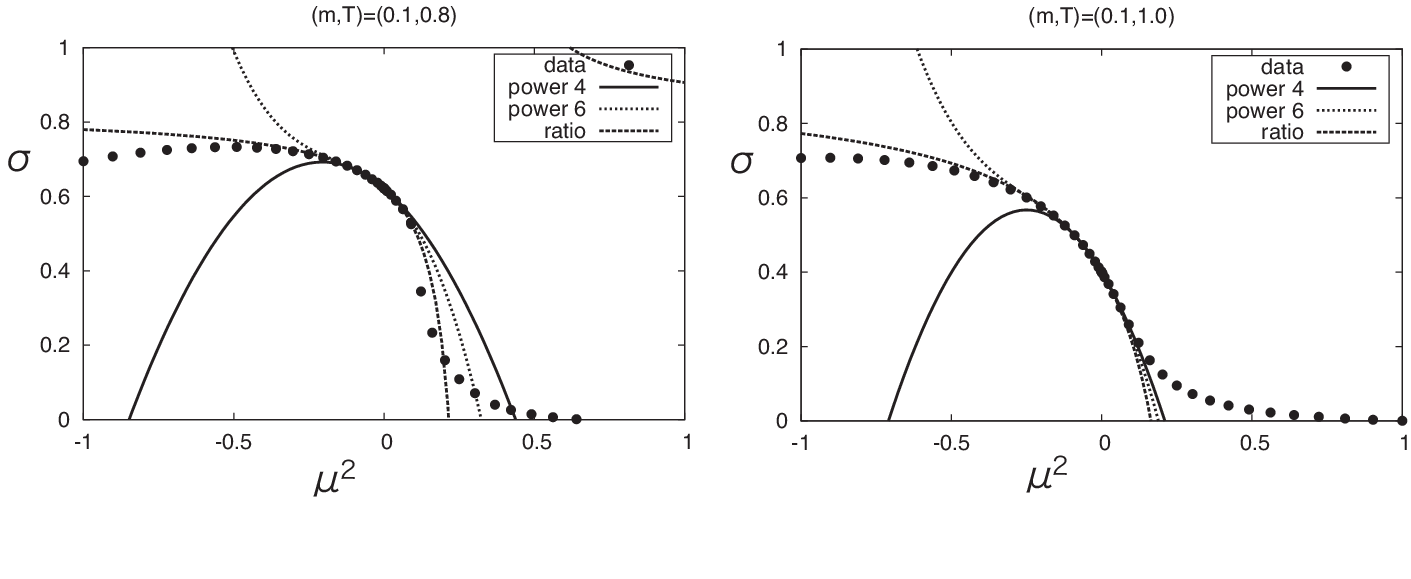}}
\caption{Fitting    at $T=0.8$ (left) and $T=1.0$ (right) for $m=0.1$. Two types of polynomial fit (the 4th and 6th degree) and ratio type are compared with the ``data".
 }
\label{fig:fit-1}
\label{fig:sigfitm01T08}
\end{figure}

\begin{figure}
%\vspace{-5mm}
        \centerline{\includegraphics[width=13cm, height=7
cm]{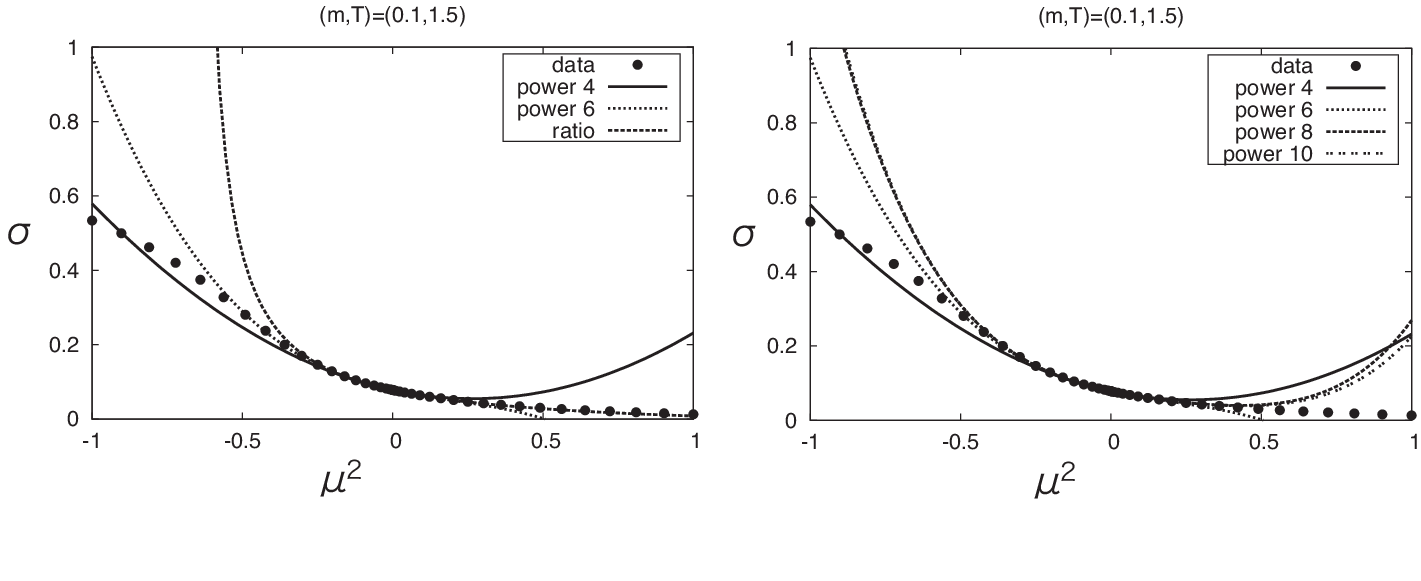}}
\caption{Fitting at   $T=1.5$ for $m=0.1$.  Two types of polynomial fit (the 4th and 6th degree) and ratio type are compared with the ``data" in the left panel.
The maximal power dependence in the polynomial fit is shown in the right panel. 
 }
\label{fig:sigfitm01T15}
\end{figure}

We divide temperature into three different regions,  $T<T_D$ (region I),   $T_D<T<T_{co}$ (region II) and  $T_{co}<T$ (region III), where $T_D$ is  the maximal temperature of the  diquark condensate phase  ($T_D=0.892$) and $T_{co}$ is  temperature of cross over at $\mu=0$ ($T_{co}=1.18$).  
It should be  noted that this classification of  the  region of temperature  are different from that in Ref.~\citen{rf:CCDP}.  Regime a and b in  Ref.~\citen{rf:CCDP} correspond to our region III, and regime c in  Ref.~\citen{rf:CCDP}  are further classified into region I and II  in our case.  In the latter, we distinguish whether analytic continuation hits   the diquark condensate phase (region I) or not (region II). 
In region I, we do not consider very low temperature because this model  incorporates only the lowest Matsubara frequencies   and is thus expected to be poor in predictability. 
The left panel in Fig.~\ref{fig:sigfitm01T08}  indicates the results of $\sigma$  for $T=0.8$. 
 At this temperature, there exists a critical point at $\mu_{c1}^2=0.089$. 
 Extrapolations to the real $\mu$ region for both of the two types of fitting are  good up to this critical point $\mu_{c1}$, but becomes extremely bad beyond $\mu_{c1}$,  as it is  expected. 
 This behavior corresponds to the behavior in the low temperature region in  Ref.~\citen{rf:CCDP}. 
When  they use a polynomial function  as an interpolation at this temperature region, the deviation between the fit and the real data occurs at the point where a chiral condensate susceptibility peak  
occurs in the real $\mu$ region.~\cite{rf:CCDP} \   In our case, this peak is observed as a sharp one at $\mu_{c1}$, where  the diquark condensate  starts to rise. In the polynomial case, if   the higher power contributions are included, the above stated feature does  not change (see Table~\ref{t:power}).\\
\begin{table}
\caption{Values of data and the results of various fits  at real values of $\mu$ for   $T=0.8$.    
 The range of used data is $0.0<\phi<0.3$, and $\Delta\phi=0.01$.    Note that a  critical point 
 exists at $\mu \simeq 0.299$.}
 \label{t:power}
\begin{center}
 \begin{tabular}{c||cccccc}\hline\hline
 {}&{$\mu=0.0$}&{$\mu=0.1$}&{$\mu=0.2$}&{$\mu=0.3$}&{$\mu=0.4$}&{$\mu=0.5$} \\ \hline
 {data}&{0.63270}&{0.61437}&{0.58886}&{0.52881}&{0.23333}&{0.10833}\\ 
 {4th degree} &{0.62178}&{0.61474}&{0.59160}&{0.54633}&{0.46888}&{0.34518} \\ 
 {6th degree} &{0.62171}&{0.61442}&{0.58957}&{0.53650}&{0.43187}&{0.23130} \\ 
 {8th degree} &{0.62170}&{0.61437}&{0.58905}&{0.53233}&{0.40771}&{0.12378} \\ 
{10th degree}&{0.62170}&{0.61437}&{0.58891}&{0.53046}&{0.39102}&{0.01637} \\  
{ratio}&{0.62169}&{0.61434}&{0.58870}&{0.52854}&{0.36526}&{-0.77936}\\
\hline
 \end{tabular}
 \end{center}
\end{table}
   In region II,  we choose $T=1.0$.  At this temperature, a cross over exists at $\mu^2=0.109$.  In the polynomial  method,  the effect  of the higher power terms becomes smaller,  and the 4th degree of   polynomial yields  as good  results as the higher degree of  polynomials  as shown in the right panel in Fig.~\ref{fig:sigfitm01T08}. Deviations from the data  become large  at a  value of $\mu$  slightly higher than the cross over point  where a chiral condensate susceptibility peak  occurs. The ratio method is also in the same quality as the polynomial ones. 
   In Table~\ref{t:T10}, we show the numerical values of the results. \\
   \begin{table}
\caption{Values of data and the results of various fits  at real values of $\mu$ for   $T=1.0$.    
 The range of used data is $0.0<\phi<0.3$, and $\Delta\phi=0.01$.    Note that a  cross over point 
 exists at $\mu \simeq 0.33$.}
 \label{t:T10}
\begin{center}
 \begin{tabular}{c||cccccc}\hline\hline
     {}&{$\mu=0.0$}&{$\mu=0.1$}&{$\mu=0.2$}&{$\mu=0.3$}&{$\mu=0.4$}&{$\mu=0.5$}\\ \hline
     {data}&{0.40000}&{0.38615}&{0.34111}&{0.25935}&{0.16310}&{0.09460}\\
     {4th degree}&{0.40007}&{0.34220}&{0.38640}&{0.25780}&{0.11710}&{-0.10244}\\
     {6th degree}&{0.40000}&{0.38611}&{0.34035}&{0.24884}&{0.08340}&{-0.20614}\\
     {8th degree}&{0.40000}&{0.38614}&{0.34060}&{0.25083}&{0.09493}&{-0.15482}\\
     {10th degree}&{0.40000}&{0.38615}&{0.34092}&{0.25511}&{0.13320}&{-0.09155}\\
     {ratio}&{0.39996}&{0.38593}&{0.33880}&{0.23825}&{0.01903}&{-0.62254}\\ \hline
 \end{tabular}
\end{center}
\end{table}
   In the higher temperature region, region III, however,  we obtained different results. 
  Figure~\ref{fig:sigfitm01T15} indicates the fitting at  $T=1.5$. In the left panel,    we compare the different types of fitting, and find that the ratio type of fit works very well.  This behavior is in good agreement with that in the higher temperature region (regime a and b) in the MC simulations.~\cite{rf:CCDP} \  In contrast to this behavior, the polynomial fit shows the behavior with very slow convergence as shown in the right panel. There, we compare the results of the polynomial fit with  the  maximal power from 4 to 10.  In Table \ref{t:T15}, we  show  the numerical values  of these fittings   at several  values of $\mu$.  In order to check if   the ratio method works well at other temperatures in region III,  we  chose    $T=1.7, 1.8, 1.9, 2.0$ and 2.2 as temperatures higher than 1.5, and found that this is indeed the case.  When we chose, on the other hand,  the value of $T$ slightly lower  than 1.5  in region III,  the same tendency is seen,   but the contrast between the ratio and polynomial types becomes less clearer  than the case of T=1.5. The results at  $T=1.3$  are listed  in Table~\ref{t:T13}. The values of the  fitting parameters used  for  $T=0.8, 1.0, 1.3$ and 1.5 are summarized in Table~\ref{t:parameters}. 
  %,  as shown in Fig.~\ref{fig:Cea1}. 
%
  \begin{table}
  \caption{Values of data and the results of various fits  at real values of $\mu$ for   $T=1.5$.    
 The range of used data is $0.0<\phi<0.3$, and $\Delta\phi=0.01$. }
 \label{t:T15}
\begin{center}
\begin{tabular}{c||cccccc}\hline\hline
{}&{$\mu=0.0$}&{$\mu=0.2$}&{$\mu=0.4$}&{$\mu=0.6$}&{$\mu=0.8$}&{$\mu=1.0$}\\ \hline
{data}&{0.07802}&{0.07142}&{0.05570}&{0.03837}&{0.02399}&{0.01353}\\ 
{4th degree}&{0.07803}&{0.07160}&{0.05861}&{0.05791}&{0.10094}&{0.23171}\\
{6th degree}&{0.07802}&{0.07141}&{0.05503}&{0.02891}&{-0.04148}&{-0.2771}\\
{8th degree}&{0.07802}&{0.07142}&{0.05578}&{0.04097}&{0.05876}&{0.26905}\\
{10th degree}&{0.07802}&{0.07142}&{0.05573}&{0.04058}&{0.05345}&{0.22534}\\
{ratio}&{0.07802}&{0.07139}&{0.05540}&{0.03713}&{0.02101}&{0.00829}\\
\hline
\end{tabular}
\end{center}
\end{table}
  \begin{table}
\caption{Values of data and the results of various fits  at real values of $\mu$ for   $T=1.3$.    
 The range of used data is $0.0<\phi<0.3$, and $\Delta\phi=0.01$. }
 \label{t:T13}
\begin{center}
 \begin{tabular}{c||cccccc}\hline\hline
 {}&{$\mu=0.0$}&{$\mu=0.2$}&{$\mu=0.4$}&{$\mu=0.6$}&{$\mu=0.8$}&{$\mu=1.0$} \\ \hline
 {data}&{0.13425}&{0.11678}&{0.07971}&{0.04646}&{0.02424}&{0.01072} \\ 
 {4th degree}&{0.13426}&{0.11721}&{0.08920}&{0.11963}&{0.32415}&{0.86470} \\ 
 {6th degree}&{0.13425}&{0.11694}&{0.08420}&{0.07916}&{0.12544}&{0.15474} \\ 
 {8th degree}&{0.13425}&{0.11671}&{0.07362}&{-0.09124}&{-1.29117}&{-7.56647} \\ 
 {10th degree}&{0.13425}&{0.11678}&{0.08174}&{0.16905}&{2.37356}&{22.91351} \\ 
 {ratio}&{0.13422}&{0.11625}&{0.07496}&{0.03110}&{-0.00495}&{-0.03183} \\ \hline
 \end{tabular}
 \end{center}
\end{table}
\begin{table}
\caption{Parameters of various fits of the chiral condensate in accordance to 
$\sigma(\phi)=A+B\phi^2+C\phi^4+D\phi^6+E\phi^8+F\phi^{10}$ (polynomial; p) or 
$\sigma(\phi)=\frac{A+B\phi^2}{1+C\phi^2}$ (ratio; r). Note that blank columns stand for terms not 
included in the fit.} 
 \label{t:parameters}
\begin{center}
\hspace*{-10mm}
\begin{tabular}{c|ccccccccc}\hline\hline
{$T$}&{Fit}&{A}&{B}&{C}&{D}&{E}&{F}&{d.o.f.}&{$\chi^2/{\rm d.o.f.}$}\\  \hline
{}&{p}&{0.62178}&{-0.68756}&{-1.67541}&{}&{}&{}&{28}&{$5.63814\times 10^{-9}$}\\ 
{}&{p}&{0.62171}&{-0.70553}&{-2.26163}&{-4.65104}&{}&{}&{27}&{$3.18334\times 10^{-11}$}\\ 
{0.8}&{p}&{0.62170}&{-0.70780}&{-2.39816}&{-7.22101}&{-14.9146}&{}&{26}&{$1.97889\times 10^{-13}$}\\ 
{}&{p}&{0.62170}&{-0.70807}&{-2.42373}&{-8.05784}&{-25.9669}&{-50.6777}&{25}&{$1.57648\times 10^{-15}$}\\ 
{}&{r}&{0.62169}&{-2.88150}&{-3.49352}&{}&{}&{}&{28}&{$4.40095\times 10^{-11}$}\\ 
\hline 
{}&{p}&{0.40007}&{-1.33932}&{-2.68272}&{}&{}&{}&{28}&{$4.65810\times 10^{-9}$}\\ 
{}&{p}&{0.40000}&{-1.35568}&{-3.21656}&{-4.23547}&{}&{}&{27}&{$8.22244\times 10^{-12}$}\\ 
{1.0}&{p}&{0.40000}&{-1.35459}&{-3.1514}&{-3.00882}&{7.11877}&{}&{26}&{$1.05268\times 10^{-12}$}\\ 
{}&{p}&{0.40000}&{-1.35399}&{-3.09274}&{-1.08930}&{32.4704}&{116.244}&{25}&{$2.02417\times 10^{-14}$}\\ 
{}&{r}&{0.39996}&{-2.43151}&{-2.66407}&{}&{}&{}&{28}&{$1.26412\times 10^{-9}$}\\ 
\hline
{}&{p}&{0.13426}&{-0.47440}&{1.20483}&{}&{}&{}&{28}&{$1.08238\times 10^{-10}$}\\ 
{}&{p}&{0.13425}&{-0.47683}&{1.12564}&{-0.62833}&{}&{}&{27}&{$6.11875\times 10^{-12}$}\\ 
{1.3}&{p}&{0.13425}&{-0.47782}&{1.06582}&{-1.75429}&{-6.53443}&{}&{26}&{$4.66163\times 10^{-14}$}\\ 
{}&{p}&{0.13425}&{-0.47769}&{1.07827}&{-1.35679}&{-1.15240}&{24.6779}&{25}&{$5.27750\times 10^{-17}$}\\ 
{}&{r}&{0.134221}&{-0.22694}&{1.91334}&{}&{}&{}&{28}&{$9.61851\times 10^{-10}$}\\
\hline 
{}&{p}&{0.07809}&{-0.17377}&{0.32745}&{}&{}&{}&{28}&{$5.2598\times 10^{-11}$}\\ 
{}&{p}&{0.07802}&{-0.17551}&{0.27069}&{-0.45033}&{}&{}&{27}&{$3.03625\times 10^{-14}$}\\ 
{1.5}&{p}&{0.07802}&{-0.17544}&{0.27492}&{-0.37034}&{0.46189}&{}&{26}&{$1.51937\times 10^{-17}$}\\ 
{}&{p}&{0.07802}&{-0.17544}&{-0.27490}&{-0.37132}&{0.45424}&{-0.03506}&{25}&{$1.57037\times 10^{-17}$}\\ 
{}&{r}&{0.07802}&{-0.05710}&{1.52002}&{}&{}&{}&{28}&{$1.93257\times 10^{-12}$}\\ 
\hline
\end{tabular}
\end{center}
\end{table}
\subsection{chiral susceptibility peak}
\begin{figure}
%\vspace{-5mm}
        \centerline{\includegraphics[width=9cm, height=7
cm]{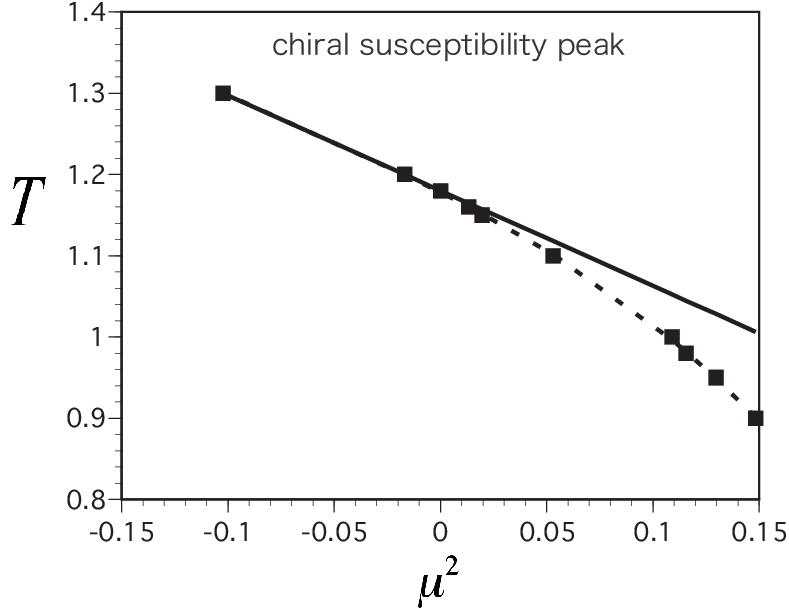}}
\caption{Locations of the chiral susceptibility peaks in $\mu^2-T$ plane for $m=0.1$, and two types of polynomial  fit: 
linear in $\mu^2$, $1.180-1.171 \mu^2$ (solid line) and the maximal power with $\mu^6$, $1.178-1.294 \mu^2-2.157 \mu^4-11.564 \mu^6$ (broken line).
 }
\label{fig:chi-sus-pk12}
\end{figure}

The analytic continuation method has been applied to the determination of the critical line.~\cite{rf:dFP} \ We here focus on the chiral susceptibility peak for this purpose. 
The chiral susceptibility $\chi_{ch}$ is calculated by  the $m$-dependence of $\sigma$. The location of  a peak of $\chi_{ch}$ in $\mu^2-T$ plane for $m=0.1$ is shown in  Fig.~\ref{fig:chi-sus-pk12}.
We fit the imaginary "data"  for $0\leq \phi\leq 0.32$ by  polynomial formulae, one is $A+B\mu^2$ and  the other $A+B\mu^2+C\mu^4+D\mu^6$,  and extrapolate them to the real $\mu$ region. These are compared with 
the results for real $\mu$ values of the model as shown in Fig.~\ref{fig:chi-sus-pk12}. 
The latter fits very well, while the former does not function at all. That higher order terms play a relevant role in the case of   chiral susceptibility peak is  in good  agreement with that  of Cea et.al. (see Fig's. 3 and 6 in the first reference in Ref.~\citen{rf:CCDP2}).

\section{Effect of higher Matsubara frequencies  }
\label{sec:sec4} 
The model we have studied so far, let us tentatively  call it the KTV model,  incorporates temperature effects only from the lowest  Matsubara frequencies.  This limitation is expected to give an influence on the behaviors of the  baryon density and chiral condensate at low temperatures, because the Matsubara frequencies form  a continuum at  zero temperature limit. 
Vanderheyden and Jackson~\cite{rf:VJ} \  introduced an extended  RMT model (VJ model) incorporating contributions from higher Matsubara frequencies.  We study this model in the imaginary $\mu$ region and investigate the analytic continuation in order to check to what extent  the results obtained in the previous section are robust, particularly, in  region III.\\ @
The partition function of  the VJ model is given by 
\begin{eqnarray}
Z=\int d\sigma d\Delta e^{-n{\tilde \Omega}}.
\end{eqnarray} 
The effective potential  is given by 
\begin{eqnarray}
{\tilde \Omega}&=&2\beta G^2(\sigma^2+\Delta^2)\nonumber-f(\sigma, \Delta), \\
f(\sigma, \Delta)&=&\sum_{\pm}\sum_{n=-\infty} ^{\infty}\ln
\left[\beta^2\left(\left(\sigma + m\pm \mu \right)^2+\Delta^2 +(2n+1)^2\pi^2T^2 \right)\right], \label{eq:actionVJ}
\end{eqnarray} 
where the summation over $n$ stands for the one over Matsubara frequencies, and the summation over $\pm$ does  their positive and negative parts, and  $\beta=1/T$. 
A term, referred to as the regular term in Ref.~\citen{rf:VJ},  is dropped in Eq.~(\ref{eq:actionVJ}), since it is independent of $\sigma$ and $\Delta$,  and irrelevant in the following argument concerning the analytic continuation.   
By defining 
\begin{eqnarray}
\Omega={\tilde \Omega}/\beta\nonumber, 
\end{eqnarray} 
the partition function reads
\begin{eqnarray}
Z=\int d\sigma d\Delta e^{-n\beta \Omega},
\end{eqnarray} 
where $\Omega$ is expressed as 
\begin{eqnarray}
 \Omega&=&2 G^2(\sigma^2+\Delta^2)\nonumber\\
 &-&\beta^{-1}\sum_{\pm}\sum_{n=-\infty} ^{\infty}\ln
\left[\beta^2\left(\left(\sigma + m\pm \mu \right)^2+\Delta^2 +(2n+1)^2\pi^2T^2 \right)\right].\label{eq:actionVJ2}
\end{eqnarray} 
If one takes only the lowest frequencies  ($n=0$ and $n=-1$) into account,  Eq.~(\ref{eq:actionVJ2}) becomes
\begin{eqnarray}
 \Omega=2 G^2(\sigma^2+\Delta^2)-2\beta^{-1}\ln
\left(\left(\sigma + m\pm \mu \right)^2+\Delta^2 +\pi^2T^2  \right) -2\beta^{-1}\ln \beta^2.
\end{eqnarray} 
In comparison to the KTV  model Eq.~(\ref{eq:action}), 
 it is noted that the definition of temperature is different from that of  the VJ model. 
Let us consider the high temperature limit, where the approximation  in  the KTV model with  only the lowest Matsubara modes incorporated  is expected to be justified.  
For this,  the  summation in Eq.~(\ref{eq:actionVJ2})  is carried out to yield~\cite{rf:VJ} 
\begin{eqnarray}
 \Omega&=&2G^2(\sigma^2+\Delta^2)-2T\sum_{\pm}\log\left(e^{E_{\pm}/2T}+e^{-E_{\pm}/2T}\right) \label{eq:actionVJ3}\\
&=&
2 G^2(\sigma^2+\Delta^2)-\sum_{\pm}\left[ E_{\pm}+2T\log\left(1+e^{-\beta E_{\pm}}\right)\right], 
\label{eq:actionVJ4}
\end{eqnarray} 
where $E_{\pm}=\sqrt{(\sigma+m\pm\mu)^2+\Delta^2}$.
By taking a  limit   $T\rightarrow \infty $ of  Eq.~(\ref{eq:actionVJ3}) and    the effective potential of the KTV model, Eq.~(\ref{eq:action}), 
we see that the VJ model is  reduced to   the KTV model at high temperatures, as expected, with the following identification of the two temperatures 
\begin{eqnarray}
4T_{\rm VJ}=T^2_{\rm KTV},
\label{eq:Temp}
\end{eqnarray} 
where $T_{\rm VJ}$ ($T_{\rm KTV}$) denotes temperature in the VJ (KTV) model.
We have checked this correspondence of  temperature by looking at the values of chiral condensates $\sigma$. 
At $T_{\rm KTV}=2.0$ $(T_{\rm VJ}=1.0)$, for example, a ratio of $\sigma$ of the VJ model to that of the KTV model is 1.030 at $\mu=0.5$ and 1.713 at $\mu=1.0$.  At $T_{\rm KTV}=6.0$ $ (T_{\rm VJ}=9.0)$, they are 1.021  ($\mu=0.5$) and 1.086 ($\mu=1.0$). 
As $T$ increases,     the difference between the VJ and the KTV models become smaller, and the range where the ratio is approximately 1.0 becomes wider near $\mu=0$.  \par
\begin{figure}
%\vspace{-5mm}
        \centerline{\includegraphics[width=9cm, height=7.5
cm]{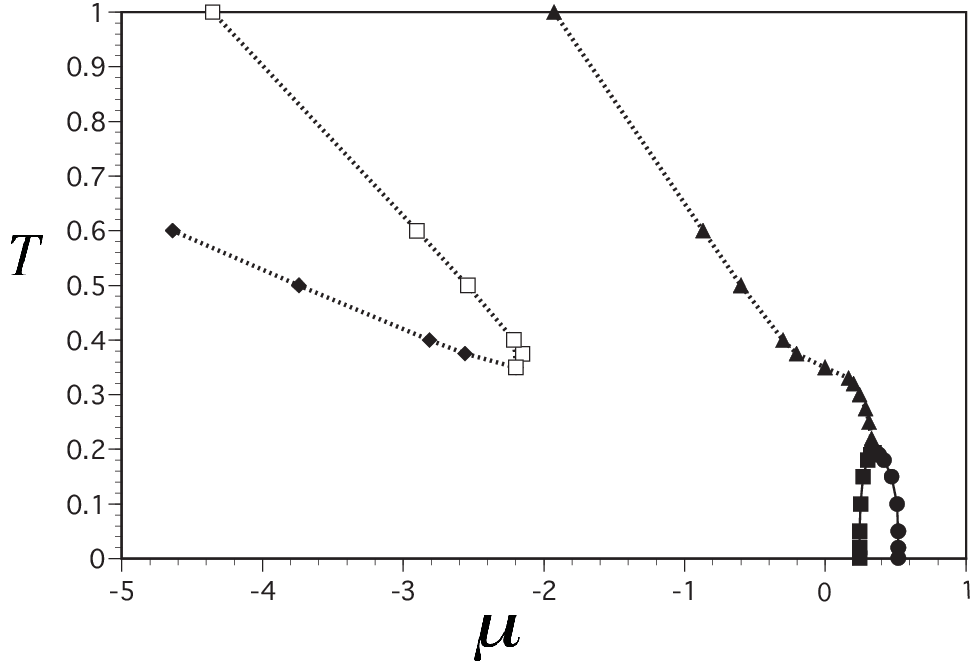}}
\caption{Phase structure in $\mu-T$ plane.  In the imaginary $\mu$ region, additional chiral susceptibility peaks appear due to the periodicity $2\pi$ in $\theta=\phi/T$. $m=0.1$. Solid  and dotted lines  indicate the second order critical line and cross over, respectively. Negative side of the horizontal axis  indicates the  imaginary chemical potential,  $i|\mu|$.   
 }
\label{fig:phase-suspeak}
\end{figure}

By numerically solving the saddle point  equations for $m=0.1$, a  similar phase structure to that in the right panel of   Fig.~\ref{fig:phasem0} is obtained  in the real $\mu$ region, but with different temperature scale.  The diquark condensate phase is located  in the region $0.245\leq \mu\leq  0.519$ at $T_{\rm VJ}=0$. The location, $\mu=0.245$,  of the lower critical point  is  in fairly good agreement with a  lowest perturbative calculation  with a small  parameter $m$,  $\mu_c^2=m/(2G^2) (\mu_c=0.223)$.   This region closes itself at $T_{\rm VJ}=0.194 (\equiv \tilde{T}_D)$.    A  cross over line starts at $(\mu, T_{\rm VJ})=(0.34, 0.194)$ and moves  towards higher $T$ and smaller $\mu$ region  as  shown   in   Fig.~\ref{fig:phase-suspeak}, crossing  the $T$-axis at  $\mu=0$ and $T_{\rm VJ}=0.35 (\equiv  \tilde{T}_{co})$.  It is noted that 
if one uses  the relation  Eq.~(\ref{eq:Temp}), these values are in good agreement with the corresponding temperature of the KTV model;  $\left(T_D\right)^2/4=0.891^2/4=0.198$, and $\left(T_{co}\right)^2/4=1.18^2/4=0.348$. 
\begin{table}
\caption{Values of data and the results of various fits at real values of $\mu$ for $T_{\rm VJ}=0.15$ in the VJ model. 
The range of used data is $0.0<\phi<0.3$ and $\Delta\phi=0.01$.  The lower (higher) critical point of the diquark condensation exists at $\mu=0.270 (0.472)$.
}
 \label{t:TVJ015}
 \begin{center}
  \begin{tabular}{c||cccccc} \hline\hline
  {}&{$\mu=0.0$}&{$\mu=0.1$}&{$\mu=0.2$}&{$\mu=0.3$}&{$\mu=0.4$}&{$\mu=0.5$} \\ \hline
  {data}&{0.47942}&{0.47414}&{0.45345}&{0.29623}&{0.10337}&{0.03243} \\
  {4th degree}&{0.47955}&{0.47473}&{0.45812}&{0.42327}&{0.35944}&{0.25159} \\
  {6th degree}&{0.47944}&{0.47426}&{0.45512}&{0.40878}&{0.30490}&{0.08376} \\
  {8th degree}&{0.47943}&{0.47417}&{0.45409}&{0.40047}&{0.25671}&{-0.13064} \\
  {10th degree}&{0.47942}&{0.47415}&{0.45384}&{0.39884}&{0.24591}&{-0.21725} \\
  {ratio}&{0.47938}&{0.47387}&{0.45062}&{0.34727}&{0.99233}&{0.63672} \\ \hline
  \end{tabular}
 \end{center}
\end{table}
\begin{table}
\caption{Values of data and the results of various fits at real values of $\mu$ for $T_{\rm VJ}=0.3$ in the VJ model. 
The range of used data is $0.0<\phi<0.3$ and $\Delta \phi=0.01$.  A susceptibility peak is located at  $\mu=0.245$.}
 \label{t:TVJ03}
 \begin{center}
  \begin{tabular}{c||cccccc} \hline\hline
  {}&{$\mu=0.0$}&{$\mu=0.1$}&{$\mu=0.2$}&{$\mu=0.3$}&{$\mu=0.4$}&{$\mu=0.5$} \\ \hline
  {data}&{0.28040}&{0.26792}&{0.23090}&{0.17619}&{0.12217}&{0.08200} \\
  {4th degree}&{0.28029}&{0.26739}&{0.22674}&{0.15250}&{0.0349}&{-0.13954} \\
  {6th degree}&{0.28038}&{0.26776}&{0.22908}&{0.16381}&{0.0775}&{-0.0853} \\
  {8th degree}&{0.28040}&{0.26790}&{0.23076}&{0.17741}&{0.1564}&{0.34229} \\
  {10th degree}&{0.28040}&{0.26792}&{0.23113}&{0.18232}&{0.20021}&{0.62457} \\
  {ratio}&{0.28026}&{0.26725}&{0.22569}&{0.14681}&{0.01016}&{-0.23318} \\ \hline
  \end{tabular}
 \end{center}
\end{table}
In the imaginary $\mu= i\phi$ region, 
the VJ model respects $2\pi$ symmetry in $\theta\equiv \phi/T$,  in contrast to the KTV model. 
This is seen,   when one sets $\Delta=0$ in the imaginary $\mu$  region, in the 
the effective potential  
 \begin{eqnarray}
\Omega=
2G^2\sigma^2-2\sum_{\pm}\log\left(\exp\left[\frac{\sigma+m}{2T}\pm i\frac{\theta}{2}\right]+\exp\left[-\frac{\sigma+m}{2T}\mp i\frac{\theta}{2}\right]\right),
% \label{eq:actionVJ4}
\end{eqnarray} 
due to the summation $\sum_{\pm}$ over positive and negative Matsubara frequencies, where $T=T_{\rm VJ}$. 
The cross over line stated above crosses the $T$-axis at $\tilde{T}_{co}$ and rises almost linearly. In addition to this,   pairs of cross over lines appear, starting at $(\phi= 2n\pi, T_{\rm VJ}=\tilde{T}_{co})$. 
In the low temperature region,  the  susceptibility peak  appeared  in the KTV model (subsction~\ref{sec:sunsec}),  although it is very broad and  low,    does not appear.  So, this structure of the peak  is  due   to the approximation limiting the Matsubara frequencies. \par
\begin{figure}
%\vspace{-5mm}
        \centerline{\includegraphics[width=8cm, height=8
cm]{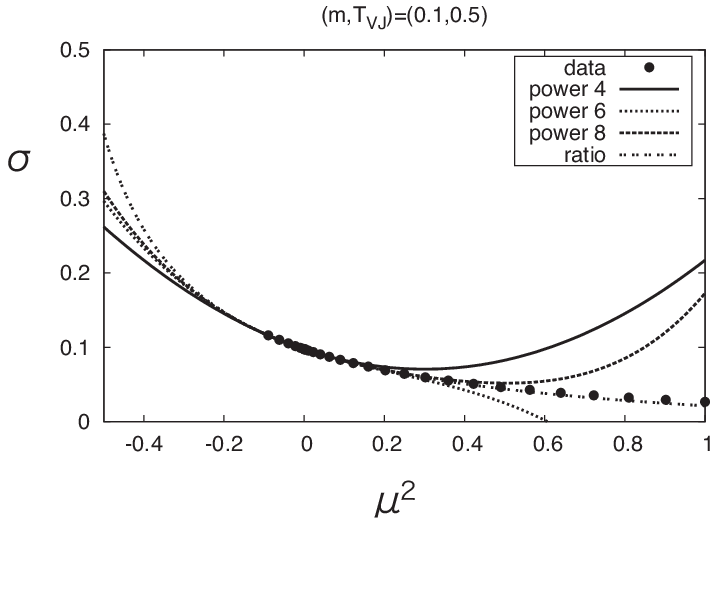}}
%VJ-fit-m01T05s.eps
\caption{Analytical continuation  of chiral condensate  at $T_{\rm VJ}=0.5$ in the VJ model  for m=0.1. 
 }
\label{fig:fit-m01T05} 
\end{figure}

 
For the purpose of the  analytic continuation,  we  studied the model by dividing   the temperature  into the  three regions as  in the previous section: $T_{\rm VJ}< \tilde{T}_D$ (region I), $\tilde{T}_D\leq T_{\rm VJ}< \tilde{T}_{co}$  (region II) and  $ \tilde{T}_{co}\leq T_{\rm VJ}$  (region III).  
At temperatures in region I,  the analytic continuation  hits the lower critical point of the diquark condensate phase at $\mu=\mu_{c1}$.  Up to this point,  both of the polynomial and ratio types of fit work but   beyond it the fits break down.   The results for  $T_{\rm VJ}=0.15$ is listed in Table~\ref{t:TVJ015}. At this temperature,  $\mu_{c1}=0.270$. At $\mu=0.25, 0.27, 0.29$, for example, the relative difference between the data and the 10th degree polynomial  (ratio) type  of  fit are 0.58\% (2.19\%),  1.48\%  (3.54\%), 23.15\%  (11.76\%). 
 In region  II,  we chose $T_{\rm VJ}=0.3$.  At this temperature, a susceptibility peak is located at  $\mu=0.245$. In the similar way to the KTV case, both of the polynomial (with degree 4 to 10) and the ratio types of fit  work well 
up to a point  slightly higher than the cross over point, but beyond this point all extrapolations rapidly deviate from the data.  
 The results for  $T_{\rm VJ}=0.3$ is listed in Table~\ref{t:TVJ03}.
  In contrast to these results,  quite different results are obtained in region III. Figure~\ref{fig:fit-m01T05} indicates the behaviors at $T_{\rm VJ}=0.5$. See also Table~\ref{t:TVJ05}.  Although the polynomial type of fits show very slow convergence, the ratio type works very well, in agreement with the results in the KTV  model in the previous section.  The truncation of the Matsubara sum does  not affect 
   the properties  of the chiral condensate at temperatures higher than the pseudo critical temperature. This fact may suggest that  it is  rather general feature in accordance with the results of MC simulations.~\cite{rf:CCDP}  \ The values of the  fitting parameters used  for all the temperatures are summarized in Table~\ref{t:parametersVJ}. \\@
\begin{table}
 \caption{Values of data and the results of various fits at real values of $\mu$ for $T_{\rm VJ}=0.5$ in the VJ model. 
 The range of used data is $0.0<\phi<0.3$ and $\Delta\phi=0.01$.}
  \label{t:TVJ05}
 \begin{center}
  \begin{tabular}{c||cccccc}\hline\hline
  {}&{$\mu=0.0$}&{$\mu=0.2$}&{$\mu=0.4$}&{$\mu=0.6$}&{$\mu=0.8$}&{$\mu=1.0$} \\ \hline
  {data}&{0.09747}&{0.09064}&{0.07403}&{0.05513}&{0.03231}&{0.02671} \\
  {4th degree}&{0.09748}&{0.09078}&{0.07644}&{0.07166}&{0.10515}&{0.21708} \\
  {6th degree}&{0.09747}&{0.09063}&{0.07356}&{0.04831}&{-0.00952}&{-0.19261} \\
  {8th degree}&{0.09747}&{0.09064}&{0.07406}&{0.05638}&{0.05750}&{0.17265} \\
  {10th degree}&{0.09747}&{0.09064}&{0.07406}&{0.05637}&{0.05740}&{0.17187} \\
  {ratio}&{0.09747}&{0.09061}&{0.07374}&{0.05391}&{0.03589}&{0.02131} \\ \hline
  \end{tabular}
 \end{center}
\end{table}
\begin{table}
 \caption{Parameters of various fits of the chiral condensate in the VJ model  in accordance to 
 $\sigma(\phi)=A+B\phi^2+C\phi^4+D\phi^6+E\phi^8+F\phi^{10}$ 
 (polynomial; p) or $\sigma(\phi)=\frac{A+B\phi^2}{1+C\phi^2}$ 
 (ratio; r). Note that blank columns stand for terms not included in the fit.}
  \begin{center}
  \label {t:parametersVJ}
   \begin{tabular}{c|ccccccccc}\hline\hline
{$T_{\rm VJ}$}&{Fit}&{A}&{B}&{C}&{D}&{E}&{F}&{d.o.f.}&{$\chi^2$/d.o.f.}\\ \hline
{}&{p}&{0.47955}&{-0.46416}&{-1.79061}&{}&{}&{}&{28}&{$1.23052\times 10^{-8}$}\\ 
{}&{p}&{0.47944}&{-0.49064}&{-2.65467}&{-6.8544}&{}&{}&{27}&{$1.27281\times 10^{-10}$}\\ 
{0.15}&{p}&{0.47943}&{-0.49517}&{-2.92689}&{-11.9795}&{-29.7373}&{}&{26}&{$1.54657\times 10^{-12}$}\\ 
{}&{p}&{0.47942}&{-0.49592}&{-2.99819}&{-14.3132}&{-60.5581}&{-141.322}&{25}&{$2.02449\times 10^{-14}$}\\ 
{}&{r}&{0.47938}&{-3.984}&{-7.24552}&{}&{}&{}&{28}&{$2.44716\times 10^{-9}$}\\ \hline
{}&{p}&{0.28029}&{-1.27399}&{-1.62144}&{}&{}&{}&{28}&{$7.7485\times 10^{-9}$}\\ 
{}&{p}&{0.28038}&{-1.25332}&{-0.94699}&{5.35106}&{}&{}&{27}&{$3.38165\times 10^{-10}$}\\ 
{0.3}&{p}&{0.28040}&{-1.24591}&{-0.50153}&{13.7362}&{48.6624}&{}&{26}&{$1.36635\times 10^{-12}$}\\ 
{}&{p}&{0.28040}&{-1.24521}&{-0.43433}&{15.9355}&{77.7099}&{133.191}&{25}&{$1.03238\times 10^{-14}$}\\ 
{}&{r}&{0.28026}&{-1.70342}&{-1.50248}&{}&{}&{}&{28}&{$1.54665\times 10^{-8}$}\\ \hline
{}&{p}&{0.09748}&{-0.17932}&{0.29892}&{}&{}&{}&{28}&{$3.40922\times 10^{-11}$}\\ 
{}&{p}&{0.09747}&{-0.18072}&{0.25322}&{-0.36259}&{}&{}&{27}&{$1.35924\times 10^{-14}$}\\ 
{0.5}&{p}&{0.09747}&{-0.18067}&{0.25605}&{-0.30932}&{0.30912}&{}&{26}&{$1.91676\times 10^{-20}$}\\ 
{}&{p}&{0.09747}&{-0.18067}&{0.25605}&{-0.30933}&{0.30898}&{-0.00063}&{25}&{$1.99022\times 10^{-20}$}\\ 
{}&{r}&{0.09747}&{-0.04686}&{1.37552}&{}&{}&{}&{28}&{$1.554\times 10^{-12}$}\\ \hline
  \end{tabular}
 \end{center}
\end{table}

\section{Summary}
\label{sec:sec5} 
We studied analytic continuation in  two-color QCD in terms of the chiral RMT with the lowest Matsubara frequencies as the temperature effect.  In each of the three different  temperature regions, we used polynomial and ratio types of fitting functions for the chiral condensate. 
In the temperature region higher than the pseudo critical point at $\mu=0$, the ratio type of  fit works well, while the power types show slow convergence.  This is the same feature as that found in the  study of MC simulations.~\cite{rf:CCDP} \   In order to check whether  this result may be affected by an approximation involved in this RMT model, 
 we also studied a RMT model incorporating all the Matsubara frequencies~\cite{rf:VJ} \ and found that  the same result  is obtained in this temperature region. 
 It might  thus  be  rather general  behavior, and  reflect   some non trivial dynamics  of the quark gluon plasma with strong correlations.~\cite{rf:DDL}  \   
 It is interesting that  study  in the imaginary $\mu$ region may  also provide useful information on the phase structure in the real $\mu$ region.~\cite{rf:SKKY} \ 
   It would then be  encouraging to  study three-color QCD in terms of RMT.   The results of this analysis  will be reported in the forthcoming paper.  We also investigated  analytic continuation of the pseudo critical line.
   Its behavior is also similar to that in the MC simulations.~\cite{rf:CCDP2} \par
In view of the phase structure,   some drawbacks of the RMT models should be  mentioned. 
The first is a tricritical point of the diquark condensate transition,~\cite{rf:KTS,rf:STV} \ which is missing 
in the RMT models.  Because of their mean field nature, the first order phase transition at higher temperature side along the critical line is not detected.  This concerns our region I, where the analytic continuation  breaks down at the critical line in any case, irrespective of the order of the transition. 
 The second  is a matter of the saturation of the interactions,  which 
 manifests as a decreasing  behavior of the diquark condensation for  $\mu>\mu_{c1}$,  as discussed  in section~\ref{sec:sec2}.  
 Temperatures in region III are outside of the region (region I) where  the saturation  concerns   in our analysis, 
 so we expect that  the above stated behaviors in region III would be little affected, or that  even if affected, it would occur  in the much larger $\mu$ region. 
The last    is  the  Roberge-Weiss symmetry,   which reflects the center symmetry of the gauge group.    This symmetry periodically  induces   critical lines at $\theta(=\phi/T)=(k+1/2)\pi$($k$: integer).
  In the RMT model,  this symmetry  is missing.  
The existence of the critical line closest to the $\mu=0$ axis   affects the analytic continuation.~\cite{rf:CCDP} \    In terms of the  imaginary $\mu$ variable,  $\phi=T\pi/2$, the distance from the $\mu=0$ axis is proportional to $T$, and  its effect becomes less serious at higher temperatures. This may be a reason why the RMT model  leads to the same conclusions as the MC simulations at higher temperature regions. 
 In the lower temperature region, on the other hand,  missing RW symmetry may cause a  discrepancy    
 between the lattice gauge theory  and the corresponding RMT model.  In fact, a use of a periodic fitting function 
 is suggested in the low temperature region in    the  study of MC simulations.~\cite{rf:CCDP} \ 
 This is, however, beyond the scope of the  present analysis, unless some RMT  model with the RW symmetry  is considered.

\section*{Acknowledgments}
We are grateful to M. Imachi, H. Kouno, M. Tachibana and A. Wipf  for useful discussion.

%\section*{Acknowledgements}
%We would like to thank ...........

%\appendix
%\section{First Appendix} %Empty argument \section{} yields `Appendix'. 
%
%\section{Second Appendix}

\end{document}